\documentclass[12pt]{article}
\usepackage{amssymb,graphicx,wrapfig}
\author{
Vladimir A. Petrov\footnote{Vladimir.Petrov@ihep.ru}
~and Nikolai P. Tkachenko\footnote{Nikolai.Tkachenko@ihep.ru}
}
\index{}
\title{On the "Flat Odderon" Model.}
\date{}

\begin{document}

\maketitle
\begin{center}
~\vspace{-13.1mm}

A.A. Logunov Institute for High Energy Physics 

NRC "Kurchatov Institute", Protvino, RF
\end{center}
\begin{abstract}
This is a critical discussion of theoretical and descriptive deficiencies of the recently
used "flat Odderon" model.
\end{abstract}
\section*{Introduction}
For more than half a century, the properties and significance of Odderon, the C-odd
partner of Pomeron and presumably asymptotically subleading w.r.t. the latter, have
been discussed with varying intensity. By now, one can find a whole spectrum of
opinions regarding the Odderon.

{\bf Experiment}:
\begin{itemize}
\item even if it exists, its influence at high energies is insignificant; 
\item the Odderon effect has already been discovered at the LHC in the region of small
         momentum transfers;
\item Odderon is unnoticeable at small transfers, but is clearly expressed in the region
         of the dip ($pp$) and shoulder (${\bar p}p$);
\item It must be sought in processes other than $pp$ and ${\bar p}p$ elastic
        scattering.
\end{itemize}

\textbf{Theory}:
\begin{itemize}
\item Odderon is the position of the rightmost singularity of the C-odd amplitude in
          the j-plane;
\item Odderon is a Regge pole.
\item Odderon is a Regge cut. 
\end{itemize}

For a brief overview of the current state of the Odderon problem, see,
for example, \cite{RYSK}, \cite{Sza} .

In our note we are more specific and would like to touch upon one extreme case of
realization of the Odderon as a "flat Regge trajectory" \cite{Rys} (actually a fixed pole)
with help of which several "two-channel" eikonal model variants were tested with a
data description on $pp$-scattering at $\sqrt{s}= 50 \div 13000$ GeV
and for $-t < 0.1 \; \mbox{GeV}^{2}$. With use of some special prescription for
calculation of the $\chi^{2}$ the authors of the paper in question claim quite a
reasonable description of the data with $\chi^{2}/ndf = 1.11$ and indicate that the
absence of the Odderon contribution worsens the quality of the description though
this contribution is small. The sign of the Odderon was also defined and the value of
the Odderon coupling $\beta_{\mathcal{O}} (t)$ at $t=0$  is estimated to be less than
that of the Pomeron. Moreover, the contribution of the Odderon to the ratio
$\Re T^{pp}(s,0)/ \Im T^{pp}(s,0)$ was found to be an order of magnitude smaller
than that proclaimed by the TOTEM collaboration.

All these problems are of significant interest and importance, which, in fact, prompted
this note.

In relation to theory we will limit ourselves to discussing only the Odderon part, while
the evaluation of the statistical part applies to the entire work as a whole.

%%%%%%%%%%%%%%%%%%%%%%%%%%%%%%%%%%%%
\newpage
%%%%%%%%%%%%%%%%%%%%%%%%%%%%%%%%%%%%

\section*{Some preliminaries}

Keeping in mind the accompanying "unitarization" (eikonal representation) and for
reference we remind the standard expression for the Odderon as a "Born amplitude"
$\mathcal{O}(s,t)$\footnote{For the sake of certainty, we will assume that we are
talking about proton-proton scattering. We do not touch here on the question of
the sign of the Odderon contribution, which is irrelevant for the subject of our
note.} defined by the negative signature Regge pole at $J=\alpha(t)$ (see e.g.
Ref. \cite{Van})
\begin{equation}
\mathcal{O}(s,t) = \pi\frac{\alpha^{'}(t)[2\alpha(t)+1]}{2\sin\left[\pi\alpha(t)\right]}
\left[ 1-e^{-i\pi \alpha(t)}\right]\Gamma_{\alpha(t)}^{2}(t)P_{\alpha(t)}(-z_{t})=
\end{equation}
\[= \pi\alpha^{'}(t)\left[\alpha(t)+\frac{1}{2}\right]\left\{ i+ \mbox{tg}
\left[\frac{\pi\alpha(t)}{2}\right]\right\} \Gamma_{\alpha(t)}^{2}(t)
P_{\alpha(t)}(-z_{t})\]
where $ \Gamma_{\alpha(t)}(t) $ is the spin $ J $ meson vertex $ \Gamma_{J}(t) $
at $ J=\alpha(t) $ and $z_{t}=1-2s/(4m_{J}^{2}-t)$. When moving to the $t$-channel,
$\Re\alpha(t)\rightarrow J = 1, 3 ,...$, and the propagator of the C-odd meson of
(generally complex) mass $ m_{J} $ is reproduced:
\begin{equation}
\mathcal{O}(s,t)\sim \frac{\Gamma_{J}^{2}(t)}{m^{2}_{J} - t} P_{J} (z_{t}).
\end{equation}
The propagator stems from $\mbox{tg}[\pi\alpha(t)/2].$
 There are no poles (physical states) in the cross-channel of amplitudes of the "alien"
(in this case positive, with even $J  $) signature.
 
This short prelude of long-known facts will be useful to fix designations in presenting
our arguments further.

\section*{Odderon in the "flat" implementation.\\ Problems in theory and the data
processing.}

\subsection*{Theory}

At "high enough" energies it is usually supposed that the  contributions of secondary
trajectories are negligibly small. In this case, it is assumed, as made also in the paper
\cite{Rys}, that only the vacuum trajectory, Pomeron, $\alpha_{\mathcal{P}}(t)$, and
its $C$-odd partner, Odderon, $ \alpha(t) $ remain. A thorough analysis of the
available data (from the ISR to the LHC) on $ pp $ and $\bar{p}p$ interactions
was undertaken in Ref.\cite{Rys} in order to assess the scale of the influence of
the Odderon exchange against the background of the Pomeron dominance. 

In relation to the Odderon the authors, following in part Ref.\cite{Lip1}, took an
extreme position, conjecturing a flat Odderon "trajectory"
\begin{equation}
\alpha(t) = 1.
\end{equation}
Direct use of "trajectory" (3) in Eq.(1) is not possible so we proceed with use of a
"regularization"
\begin{equation}
\alpha(t) \rightarrow  1 + \alpha^{'}t
\end{equation}
at  low $ t $ and then monitor the consequences.  
Taking afterwards $ \alpha{'}= 0 $ we reproduce the "flat Odderon" premise
of Ref. \cite{Rys}. Substiting $\alpha (t) \rightarrow  1 + \alpha{'}t$ in Eq.(1) and
considering it in the limit $\alpha{'} \rightarrow 0$, $t\rightarrow 0$ we get 
\begin{equation}
\mathcal{O}(s,t) \sim  
\frac{\Gamma_{1}^{2}(t)P_{1}(z_{t})}{-t}
\end{equation}
Equation (5) means (cf.Eq.(2)) nothing more than the presence of a massless vector
hadron in the $p$-wave partial amplitude of the $t$-channel since in this case the
amplitude of the negative signature is physical (odd angular momentum, $J=1$). 

As far as is known, such hadrons have not been observed and their appearance can be
avoided only by  making the the vertex $\Gamma_{1}^{2}(t)$ vanishing at $t = 0$.
Moreover, due to analyticity of $ \Gamma_{1}(t) $, zeroing can only be of the type
$\Gamma_{1}^{2} (t) \rightarrow$ const $t^{2N}$ where $N\geq 1$ is an integer .

So we see that at the level of the Born amplitude the Odderon decouples from the
proton at $t= 0$. The real part decouples no weaker than $\sim t$, while the
imaginary part decouples  no weaker than $\sim{t}^{2}$.

Of course, what was said above about the massless pole applies only to the Born term,
but constructing the full eikonal series requires integrating the products of the Born
terms over the momentum transfers $t_{i}$ including $t_{i}=0$ , which leads to
logarithmic divergences. The trick with $\Gamma_{1}^{2}(t)\sim t^{2}$ could "save"
the situation, but further discussion of this option lies beyond our prerogatives.

\subsection*{The data description in the "flat" Odderon scenario}

Leaving aside criticism of the purely theoretical foundations of the Odderon part of the
paper \cite{Rys}, let us nevertheless consider its descriptive part in relation to the
available data on elastic pp scattering.

However, first, let us make one important remark. There is extensive literature on
the analysis of diffraction scattering parameters at various energies. However, in
this paper, we rely primarily on experimental data at 13 TeV for the following reason:
in our view, the measurements at 13 TeV are unique in that they were first
performed on the same beam and at the same energy using two different
instruments~-- ATLAS and TOTEM. These data turned out to differ significantly. As
a consequence of this fact, the extracted theoretical parameters differ significantly,
and the authors of numerous papers point to a variety of reasons for their
discrepancies. For example, in \cite{recensent1}, the authors point to an
underestimation of systematic errors at the TOTEM setup and the possibility of a
slower increase in the total cross section in the pp interaction. A significant caveat
is noted in the Pierre Auger Observatory report \cite{Pierre Auger} which
emphasizes the preference for the ATLAS total cross-section data over the
TOTEM data.

There are numerous such papers examining the possible causes of these different
experimental data (see, e.g. \cite{Gra}). However, the authors of the ATLAS and
TOTEM experimental setups remain silent on the reasons for such divergent
measurements of the same differen tial cross-section at 13 TeV. A joint analysis of
the sources of these different results, which are a result of the design of these
experimental setups, would be fundamentally important. As a consequence of this
state of affairs, we are forced to rely solely on published experimental data from
these setups, which significantly contradict each other. And what to do with these
contradictions in the absence of an analysis by their creators is anyone's guess.

There is no doubt that these measurements provide approximately correct values for
the parameters of various theoretical models, but they are fundamentally unsuitable
for sifting through different theoretical concepts (calibrating theories). For this
reason, we rely only on published experimental results from the ATLAS and TOTEM
facilities at 13 TeV and do not delve into the details of their differences.

The obvious, significant differences in experimental measurements at 13 TeV,
identified for the first time, also give reason to doubt the measurements at other
energies. In essence, this assertion is the main content of our note. For this
reason, we do not analyze the sources of the various measurement results, of
which there are already a large number.

It is important to note that authors provide both the central values of the fitted
parameters of the models and their errors. However, the parameters extracted from
the models (for example, the parameter $\rho$ \cite{add} ) are provided only with
their central values; for some reason, the authors do not provide the errors which
can be of a paramount importance for the final conclusions. Namely, in the
experimental data at the $t$ intervals the values of $|t|$ were carefully selected
based on the high confidence level (at least $80\%$) for the extracted parameters.
In particular, the work \cite{Tk} shows that $\rho = 0.1 \pm 0.04$, i.e. the extracted
value is determined with an accuracy of $40\%$ (p. 054003-9). This is fundamentally
at odds with the error of the authors of the TOTEM experiment, who groundlessly
claim an error in the $ \rho{-}$parameter to be only $10\%$ (with the same central
value). It is clear that with such an error value, it is at least premature to talk about
a discrepancy with the predictions of the COMPETE collaboration\cite{COMP}. So,
arguing about the difference in the values of the parameters in this work is
generally pointless without indication of the errors. The authors of \cite{Rys} refer to
the work just mentioned \cite{Tk}, but they cite intermediate values of  the
parameter $\rho$ from this work, and not its final value and its error from the
conclusions of the referenced article. Moreover, the authors, referring to work\cite{Tk},
fairly note that its results were obtained on the basis of the TOTEM experiment only.
Just because the official results of the ATLAS experiment did not yet exist at
that time. 

As soon as the latter appeared, they were studied with the same care as in work
\cite{Tk}. These results are presented, for example, in Ref. \cite {TkA} , to which the
authors of \cite{Rys} also refer. The conclusions of this work show that the extracted
parameters $ \sigma_{tot}$ and $\rho$ have significantly lower values than in the
TOTEM experiment. Moreover, the parameter $\rho$ itself changes significantly
depending on the working array of experimental parameters used to extract it (always
at a confidence level significantly higher than $50\%$) and thus even in a single
ATLAS experiment it is extracted significantly ambiguously. In any case, it is essentially
lower than that in the TOTEM experiment. 

Thus, these two experiments contradict each other on the essence. Either one of
them is incorrect, or they are both incorrect.

\begin{wrapfigure}[18]{l}{90mm}
\begin{tabular}{|l||c|c|c|c|c|} \hline
\multicolumn{6}{|l|}{{\bf Table III}} \\ \hline
\multicolumn{6}{|c|}{ Ensemle A}  \\ \hline
DoF ($\ nu$) & 332 & 332 & 332 & 332 & 332 \\ \hline
$\chi^2  / \nu$ & 0.96 & 0.97 & 0.97 & 0.97 & 0.96 \\ \hline
{\bf CL [\%]} & {\bf 69.2} & {\bf 64.4} & {\bf 64.4} &
{\bf 64.4} & {\bf 69.2} \\ \hline \hline
\multicolumn{6}{|c|}{ Ensemle T}  \\ \hline
DoF ($\nu$) & 418  & 418  & 418 & 418 & 418    \\ \hline
$\chi^2 / \nu$ & 1.28 & 1.30 & 1.29 & 1.28 & 1.27 \\ \hline
{\bf CL [\%]}$\cdot 10^3$ & {\bf 6.8} & {\bf 2.3} & {\bf 4.0} &
{\bf 6.8} & {\bf 11.3} \\ \hline \hline
\multicolumn{6}{|c|}{ Ensemle A $\oplus$  T}  \\ \hline
DoF ($\nu$)      & 504  & 504 & 504  & 504 & 504   \\ \hline
$\chi^2 / \nu$ & 1.11 & 1.12 & 1.11 & 1.10 & 1.09 \\ \hline
{\bf CL [\%]}  &{\bf 4.3}&{\bf 3.1}&{\bf 4.3}&{\bf 5.9}&
{\bf 7.9}\\ \hline
\end{tabular}
\caption{The values of the confidence levels in Table III (see \cite{Rys}).}
\label{TableIII}
\end{wrapfigure}

The inconsistency of the differential cross-section measurements in these experiments
is already evident from the sets of their experimental data. The distance between the
central points of these experiments (at different values of  $t$) can reach $2.5$ full
standard deviations of the TOTEM and more than $ 8 $ full standard deviations of the
ATLAS. In either case, these are unacceptably large values to present the same physical
quantity.

The ambiguity of measurements of differential cross sections in two different
experimental facilities is also observed at $\sqrt{s} =7 $ and $ 8 $ TeV, which means
that the extracted values would differ at these energies. Moreover, this ambiguity is
most pronounced at $\sqrt{s}= 13$ TeV. The results of measurements of the same
physical quantity at the LHC at the same energies in two different experiments show
that even at lower energies, where there were no two devices measuring differential
cross sections, their values could differ from the true values. For this reason, all joint
fits at different energies should always be carried out, for example, using some factors
for the experimental data, their own for each experiment, and which are considered as
fit parameters. This would allow artificially reducing very different experimental data
to a more or less uniform array.

In fact, the authors of this work do exactly that, although they attribute the fitted
multipliers not to experimental data, but to a theoretical function for differential
cross-sections. At the same time, they did not bother to explain how this operation will
(or will not) be reflected in the extracted parameters (and their errors). It seems to us
that it would be correct to be concerned with the accuracy and correctness of the very
experimental measurements, rather than engage in arithmetic exercises in describing
unacceptably different measurements of the same physical parameters, and therefore
incorrectly representing them.

Another remark to the description of the results of this work. Its authors provide the
extracted parameters, adding to them data on the number of degrees of freedom and
the value of the total $\chi^{2}/pdf$. 

%%%%%%%%%%%%%%%%%%%%%%%%%%%%%%%%%%%%%
%%%%%%%%%%%%%%%%%%%%%%%%%%%%%%%%%%%%%
%%%%%%%%%%%%%%%%%%%%%%%%%%%%%%%%%%%%%
These results from the authors' fittings \cite{Rys} allow us to calculate the confidence
level of their parameter extraction results. We have supplemented Table III of the
authors' table with rows detailing the confidence levels of these results~-- see Fig.
\ref{TableIII}.

The supplemented table shows that supplementing the general data set with
ATLAS data at 13 TeV yields a decent confidence level—over 50%
(Ensemple A). Meanwhile, a similar supplement with TOTEM data at 13 TeV
reduces the confidence level to practically zero (Ensemple T). Clearly, drawing conclusions after this that the theoretical parameters extracted from one
array are preferable to the same parameters extracted from another
array is pointless. Against this background, claims that "the presence of a C-odd
(odderon) contribution significantly improves" the quality of the data description appear speculative and unfounded. And even more so, combining the overall data set
by simultaneously adding data from the ATLAS and TOTEM experiments at 13 TeV
(Ensemle A $\oplus$ T) is also an exercise devoid of any meaning.

%%%%%%%%%%%%%%%%%%%%%%%%%%%%%%%%%%%%%
%%%%%%%%%%%%%%%%%%%%%%%%%%%%%%%%%%%%%
%%%%%%%%%%%%%%%%%%%%%%%%%%%%%%%%%%%%%

\section*{Conclusions}

Summarizing all the above said, we have to conclude that, unfortunately, both the
theoretical basis and the statistical arguments in the paper \cite{Rys} cannot be
considered sound. Accordingly, the same goes for their physical findings.

We hope that our remarks could be helpful for improving the further studies in this
field.


\begin{thebibliography}{99}

\bibitem {RYSK}

Mikhail G. Ryskin,

\textit{Current Status of the Odderon.}

Talk given at the Conference “Hadron Structure and Fundamental Interactions: from
Low to High Energies", Gatchina, Russia, July 8 – 12, 2024 and XXXVI Interna-tional
Workshop on High Energy Physics “Strong Interactions: Experiment, Theory,
Phenomenology”, Protvino, Russia,

e-Print: 2408.01990 [hep-ph]


\bibitem {Sza}
 
 I. Szanyi and T. Csörgő,
 
\textit{The ReBB model and its H(x) scaling version at 8 TeV: Odderon exchange is a certainty.}

Eur.Phys.J.C 82 (2022) 9, 827.


%%%%%%%%%%%%%%%%%%%%%%%%%%%%%%%%%%%
% 1
%%%%%%%%%%%%%%%%%%%%%%%%%%%%%%%%%%%
\bibitem{Rys}
E.G.S. Luna, M.G. Ryskin, V.A. Khoze,

\textit{Odderon contribution in light of the LHC low-$t$ data.}

Phys.Rev.D 110 (2024) 1, 014002; 

\bibitem {Van}

L. Van Hove,

\textit{Regge pole and single particle exchange mechanisms in high energy
collisions.}

Phys.Lett. 24 (1967) 183-184.

\bibitem {Lip1}
J. Bartels, L.N. Lipatov, G.P. Vacca,

\textit{A New odderon solution in perturbative QCD}.

Phys.Lett.B 477 (2000) 178-186 

\bibitem {Tk}
Vladimir A. Petrov and Nikolai P. Tkachenko,

\textit{Coulomb-nuclear interference: Theory and practice for $ pp $
-scattering at 13 TeV}.

Phys.Rev.D 106 (2022) 5, 054003.

\bibitem {add}

Although not directly related to the subject of discussion, we note one important
circumstance, in most cases ignored. The fact is that the extraction of the parameter
$\rho$ from the data is fundamentally dependent on the model used for the amplitude
of the purely strong interaction. This was mentioned, e.g. in the paper

A. Donnachie and  P.V. Landshoff,

\textit{Lack of evidence for an odderon at small t.}

Phys. Lett.B 831 (2022) 137199. 

\bibitem {COMP}
B. Nicolescu, J.R. Cudell, V.V. Ezhela, P. Gauron, K. Kang,Yu.V. Kuyanov, \\
S.B. Lugovsky, E. Martynov, E.A. Razuvaev, N.P. Tkachenko,  

\textit{Analytic amplitudes for hadronic forward scattering: COMPETE update}.

Nucl.Phys.B Proc.Suppl. 117 (2003) 400. 

\bibitem {TkA}

Vladimir A. Petrov and  Nikolai P. Tkachenko,

\textit{TOTEM-ATLAS ambiguity: Shouldn't one worry?}.

 Nucl.Phys.A 1042 (2024) 122807.
 
\bibitem{Pierre Auger}
 
\noindent {\small https://indico.cern.ch/event/1371434/contributions/6667415/attachments/3143832/5580542/ISMD2025{\_} xsec{\_}tkachenko.pdf}

\bibitem {Gra}
Per Grafström and Rafał Staszewski,

\textit{Extraction of low-mass diffractive cross section from the discrepancy between ATLAS and TOTEM total cross sections.}

Eur.Phys.J.C 85 (2025) 8, 873 

\bibitem{recensent1}

Eur.Phys.J.C 85 (2025) 8, 873

https://arxiv.org/pdf/2502.13618

\end{thebibliography}
\end{document}